\documentclass{article}
\usepackage{hiph-preprint}
\usepackage{epsfig}
\usepackage{graphicx}
\usepackage{dcolumn}
\usepackage{bm}

\volnumber{22} \issuenumber{10} \edyear{2005}                            
\frompage{000} \topage{000}                                              
\recrevdate{15 October 2005}                                             
\def\rightmark{The new computer program for relativistic hydrodynamical model}
\def\leftmark{D. Kikola, W. Peryt, Y. M. Sinyukov, M. Slodkowski, M. Szuba}
 
\title{The new computer program for three dimensional 
relativistic hydrodynamical model} 
\authors{ 
{D. Kikola$^1$, W. Peryt$^1$, Y. M. Sinyukov$^2$, M. Slodkowski$^1$, M. Szuba$^1$  
\index{Kikola, D.} 
\index{Slodkowski, M.} 
}\\[2.812mm]
{\normalsize
\hspace*{-8pt}$^1$ Faculty of Physics, Warsaw University of Technology,\\ 
00-662 Warsaw Koszykowa 75, Poland\\[0.2ex] 
\hspace*{-8pt}$^2$ Bogolyubov Inst Theor Phys \\  
03143 Kiev Metrologichna 14b, Ukraine
}}
 
\abstract{An effective computer program for three dimensional relativistic hydrodynamical 
model has been developed. It implements a new approach to the early hot phase of 
relativistic heavy-ion collisions. The computer program simulates time-space 
evolution of nuclear matter in terms of ideal-fluid dynamics. 
Equations of motions of hydrodynamics are solved making use of finite difference methods.
Commonly-used  algorithms of numerical relativistic hydrodynamics  
RHLLE and MUSTA-FORCE have been applied in simulations. 
To speed-up calculations, parallel processing has been made available for solving hydrodynamical equations. The test results of 
simulations for 3D, 2D and Bjorken expansion are reported in this paper. 
As a next step we plan to implement the hadronization  algorithm  by 
implementing the continuous particle emission for freeze-out and comparing it with  Cooper-Frye formula.}

\PACS{see: http://www.aip.org/pacs/ }
 
\makeindex
\begin{document}
 
\maketitle

\section{Introduction}\label{intro}
The hydrodynamic approach to multiparticle production 
assumes  the initial state a very hot and dense strongly interacting 
matter soon after the collision, which then expands hydrodynamically  and 
lasts until the stage when the picture of continuous medium has cased to be valid. 
That stage, so-called freeze-out, is typically described by the so-called Cooper-Frye 
prescription (CFp) that treats the system at the decay stage of evolution 
as a locally equilibrated ideal gas at some hypersurface. However, 
this prescription produces some serious problems if the hypersurface 
is, at least partially, not space-like, and - what is even more important - 
the results of many studies based on cascade (transport) models and experimental data contradict 
the idea of sudden freeze-out. The particles escape from the system during  
the whole period of its evolution and do not exhibit signs of a local equilibration at its late stages. Attempts to overcome the problem 
is made by hybrid hybrid "hydro" + "cascade" models. \cite{bib1}

\section{The relativistic hydrodynamic equations}\label{details} 

After the first  stage of a heavy ion collision, state of matter can be described in terms of fluid dynamic. 
Evolution of fluid dynamics is simulated using finite difference methods. 
Resolving differential equations on 3D lattice of physical cells represents 
time-space evolution of matter.  
The relativistic hydrodynamics equations represent energy-momentum and charge conservation.
For heavy ion collisions, the conserved charge is e.g. the (net) baryon number, 
the (net) strangeness etc. Provided that matter is in  local thermodynamical 
equilibrium, the energy-momentum tensor and the charge current assume ideal fluid. \cite{bib2}

\vspace*{-15pt}
\begin{eqnarray}
\partial_{\mu}T^{\mu\nu}=0
\end{eqnarray}
\vspace*{-25pt}
\begin{eqnarray}
\partial_{\mu}N^{\mu}=0
\end{eqnarray}

The equations of ideal fluid-dynamics are completed by specifying an equation of state (EoS)
for the matter under consideration in the form p = p(e,n)

The laboratory quantities {\bf  R} ({\it mass density}),{\bf M} ({\it momentum density}), and {\bf E} ({\it energy density})  
are related to the quantities in the local rest frame: e (energy density) and n (mass density) 
and to the fluid velocity v thrught a set of non-linear transformations.
\begin{eqnarray}
\partial N\equiv\partial_{t}R+\bigtriangledown\cdot\left(Rv\right)=0
\end{eqnarray}
\vspace*{-25pt}
\begin{eqnarray}
\partial T^{\mu\nu}\equiv\partial_{t}E+\bigtriangledown\cdot\left[\left(E+p\right)v\right]=0
\end{eqnarray}
\vspace*{-25pt}
\begin{eqnarray}
\partial T^{\mu i}\equiv\partial_{t}M+\bigtriangledown\cdot\left(M^{i}v\right)+\delta_{i}p=0
\end{eqnarray}

The ideal gas equation of state (EoS): $p=\left(\gamma-1\right)\left(\epsilon-n\right)$
and its ultrarelativistic limit ($\epsilon \gg n$): $p=\left(\gamma-1\right)$
with the adiabatic index $1\leq\Gamma\leq2$ \cite{bib3}

\section{Ellipsoidal flow}\label{intro}

In \cite{bib4} a new class of analytic solutions of the relativistic hydrodynamics was proposed 
for 3D asymmetric flows at soft EoS p=const. In general this type of solution describes expansion with the ellipsoidal flow. If one defines the
initial conditions on the hypersurface of constant time, say
$t=0$, then $t$ is a natural parameter of the evolution. Such a
representation of the solutions similar to the Bjorken and Hubble
ones with velocity field $v_{i}=a_{i}x_{i}/t$ has property of an
infinite  velocity increase at $x\rightarrow\infty$. A real fluid,
therefore, can occupy only the space-time region where
$|\textbf{v}|<1$. In \cite{bib4} a  solution of the relativistic hydrodynamics was proposed 
for anisotropic expansion of finite system.
Proper choice of parameters in such solution induces formally Hubble-like velocity profile.
This new class of analytic solutions for 3D relativistic expansion with anisotropic 
flows can describe the relativistic expansion of finite systems towards vacuum. 
They can be utilized for a description of the matter evolution in central and non-central 
ultra-relativistic heavy ion collisions, especially during deconfinement phase 
transition and the final stage of evolution of hadron systems. Also, the solutions 
can serve as a test for numerical codes describing 3D asymmetric flows in the 
relativistic hydrodynamics.

\section{Algorithms for numerical hydrodynamics}\label{intro}
Solving three dimensional hydrodynamics equations is done by 
using operator splitting method - by solving a sequence 
of one-dimensional equations of type:
\begin{eqnarray}
\partial_{t}U+\partial_{x}F\left(U\right)=0
\end{eqnarray}
using finite differences scheme: 
\begin{eqnarray}
U_{i}^{n+1}=U_{i}^{n}-\frac{\Delta t}{\Delta x}\left[F\left(U_{i+1/2}^{n}\right)-F\left(U_{i-1/2}^{n}\right)\right]
\end{eqnarray}
where: $F\left(U_{i+1/2}^{n}\right)$, $F\left(U_{i-1/2}^{n}\right)$ - intercell numerical fluxes.

\begin{figure}[ta]
\vspace*{-5pt}
\begin{center}
{$$\psfig{figure=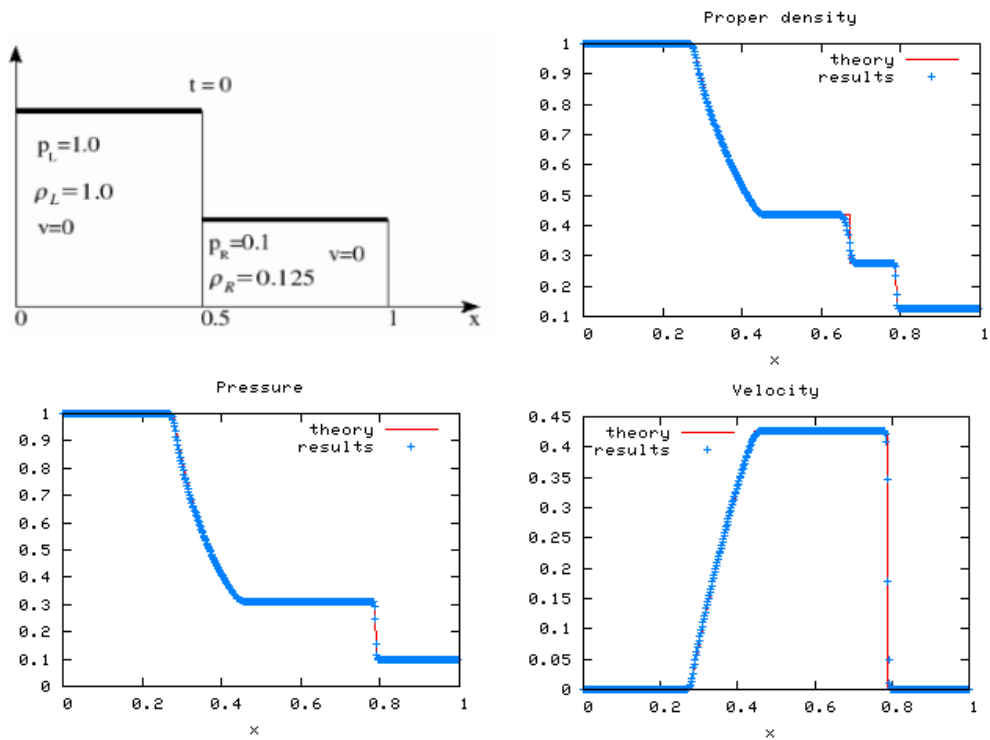}$$}
\end{center}
\vspace*{-20pt}
\caption{\label{fig2}
Solution of shock tube problem for x/t=0.4: Musta-Force flux, $\Delta$x = 0.02, $\Delta$t = $\Delta$x/4, grid of 500 zones, number of time steps: 800, EoS: ideal gas with an adiabatic index $\Gamma$ =1.4, arbitrary units.}
\end{figure}

\begin{figure}[tb]
\vspace*{-5pt}
{$$\psfig{figure=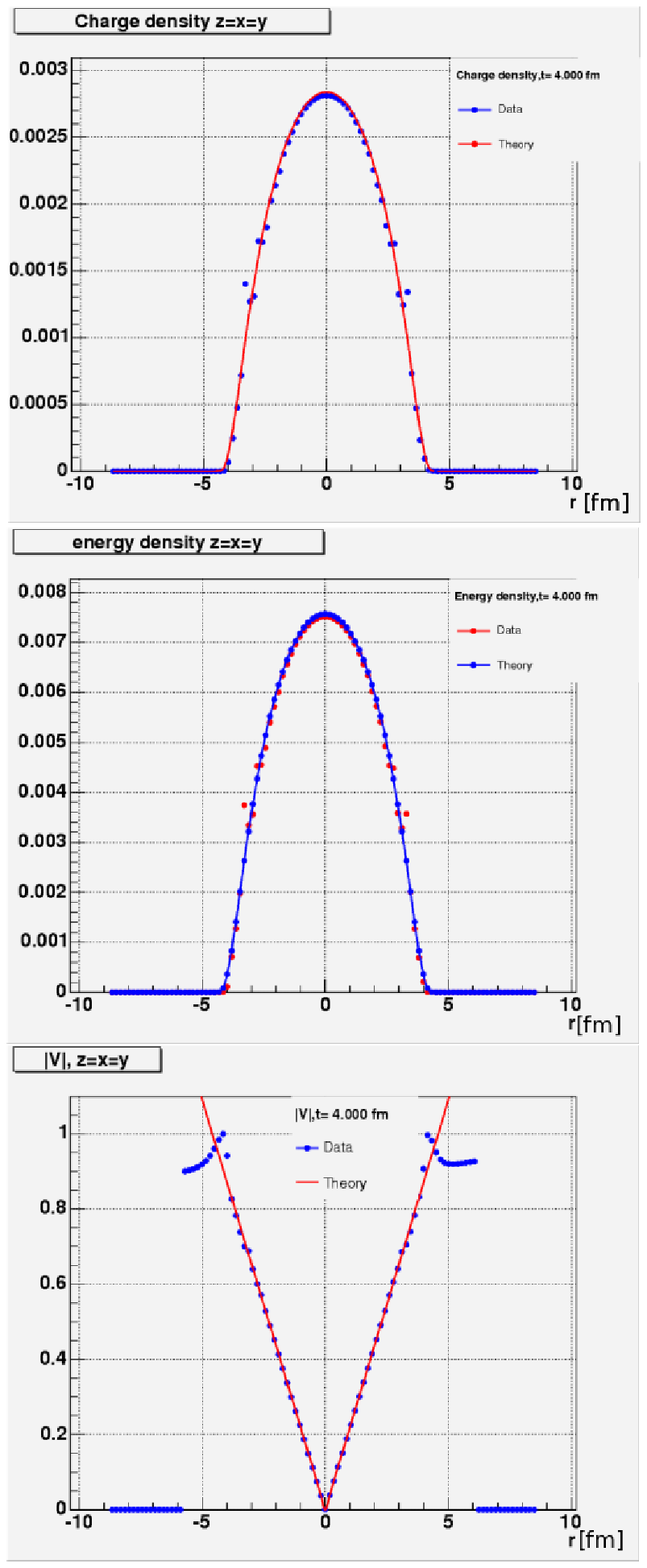}$$}
\vspace*{-20pt}
\caption{\label{fig4}
The comparison of exact and numerical solutions (using Musta-Force flux) for ellipsoidal flow with constant pressure p=0 for time t=4 fm/c, arbitrary units. Non-zero value of  velocity for  $|r|>4.5.$ fm (where the maximal  range of matter defined by velocity of light is 4.5 fm)  is caused by numerical dissipation of the matter. }
\end{figure}

\subsection{Numerical fluxs}\label{details}

\subsubsection{HLLE}\label{details}

HLLE (Harten-Lax-van Leer-Einfeldt)\cite{bib3} is a  Godunov-type algorithm, i.e. it does not apply the full solution 
of the Riemann problem but approximates it by a region of constant flow. 
In general, the good accuracy of Godunov-type fluxes results from the opening 
of the Riemann fan and picking up a single value at cell interface.
Complete (exact or approximate) Riemann solvers recognize all waves in the Riemann 
fan and therefore provide good resolution of delicate features of the flow, such as 
contact discontinuities. Incomplete Riemann solvers (e.g. HLLE flux) do not recognize 
the intermediate waves in the Riemann fun and lump them all in one (averaged) state,.
The HLLE method is very efficient and it is exact for single shocks, but can be very 
dissipative -  especially at contact discontinuities, what can be important during simulations of expansion of matter towards vacuum. 

\subsubsection{Musta-Force}\label{details} 

A very simple and general approach to the construction of numerical fluxes, which combines 
the simplicity of centered Fluxes and the good accuracy of the Godunov method, is the 
Multi-Stage (MUSTA) approach\cite{bib5}. The MUSTA approach develops upwind numerical 
fluxes by utilizing centered fluxes (FORCE flux in our case) in a multi-stage 
predictor-corrector fashion. Effectively, MUSTA can be regarded as an approximate 
Riemann solver in which the predictor step opens the Riemann fan (self-similar solution) 
and the corrector step makes use of the information extracted from the opened Riemann fan. 
This is precisely the information needed for the upwind numerical flux.
The key idea of the original MUSTA is to open the Riemann fan by solving the local Riemann 
problem numerically rather than analytically, i.e. solving it by evolving 
the initial state in time via the governing equations. It denotes that we do not explicitly make use of 
wave propagation information in the construction of the numerical flux.

\section{Parallel processing solutions for solving hydro equations}\label{techno}  

Even with optimised code, complete three-dimensional calculation 
of hydrodynamic is an extremely time-consuming task. 
In order to mitigate this problem, attempts are being made to take 
advantage of properties of numeric algorithms we use - namely, 
the fact they only operate on a fragment of computation space at a time 
- and increase simulation speed through parallel computation. 
Two different approaches are being evaluated and tested independently 
to obtain optimal results.

\subsection{PVMFlower project}\label{techno}  

Takes advantage of the Parallel Virtual Machine framework, 
an established solution for distributed computing. Source with more information: http://www.csm.ornl.gov/pvm/

\subsection{Hydrogrid project}\label{techno}  

This is a scalable network computers solution for hydro program 
simulation, based on TCP/IP network protocol. 
This project consists only of standard Unix/Linux communication components.
 
\section*{Acknowledgments}

The research described in this publication was made possible in part  by
Award No. UKP1-2613-KV-04 of the U.S.Civilian Research ${\&}$ Development
Foundation for the Independent States of the Former Soviet Union (CRDF) and Warsaw University of Technology Grant No. 503G 1050 0018 000. Research carried out within
the scope of the ERG (GDRE): Heavy ions at ultrarelativistic energies - a European Research Group comprising IN2P3/CNRS, Ecole des Mines de Nantes, Universite de Nantes, Warsaw University of Technology, JINR Dubna, ITEP Moscow and Bogolyubov Institute for Theoretical Physics NAS of Ukraine.

\vfill\eject

\begin{thebibliography}{99}  
  
\bibitem{bib1}Yu.M. Sinyukov, S.V. Akkelin, Y. Hama: "On freeze-out problem in hydro-kinetic  approach to A+A collisions", {\it Phys.Rev. 
    Lett.} {\bf 89 } (2002) 052301.  
 

\bibitem{bib2}D.H. Rischke, S. Bernard, J.A. Maruhn: "Relativistic Hydrodynamics for Heavy-Ion Collisions - I. General Aspects and Expansion into Vacuum", 
    {\it nucl-th/9504018}.
 

\bibitem{bib3}D.H. Rischke, "Fluid Dynamics for Relativistic Nuclear Collisions", {\it nucl-th/9809044}.

  
\bibitem{bib4}Yu.M. Sinyukov, Iu.A. Karpenko: "Quasi-inertial ellipsoidal flows in relativistic hydrodynamics" ,ucl-th/0505041; "Ellipsoidal flows in relativistic hydrodynamics of finite systems",
arXiv:  {\it nucl-th/0506002} (to be published in Acta
Phys. Hung. (2006)).
 
\bibitem{bib5}E F Toro: "Multi-Stage Predictor-Corrector Fluxes for Hyperbolic Equations", {\it NI03037-NPA }


\end{thebibliography}
\end{document}